\documentclass[twocolumn,prl,showkeys,amsmath,amssymb]{revtex4}

\usepackage{graphicx}
\usepackage{dcolumn}
\usepackage{bm,color}

\begin{document}

\preprint{APS/}

\title{Zero Sound Propagation in Femto-Scale Quantum Liquids}

\author{Yoritaka Iwata}
 \email{y.iwata@gsi.de}
 \affiliation{GSI Helmholtzzentrum f\"ur Schwerionenforschung, Darmstadt, Germany}

\date{\today}

\begin{abstract}
Charge equilibration has been recognized as a dominant process at the early stage of low-energy heavy-ion reactions.
The production of exotic nuclei is suppressed under the appearance of charge equilibration, in which the proton-neutron ratios of the final reaction products are inevitably averaged. 
Therefore charge equilibration plays one of the most crucial roles in the synthesis of chemical elements. 
Focusing on how and when the charge equilibration takes place, the zero-sound propagation in femto-scale quantum liquids is explained.
\end{abstract}

\maketitle

\section{Introduction}
This review article is concerned with Fermionic property of heavy ions (many-nucleon systems with the size up to several 10-femtometers) colliding at the energy of a few MeV per nucleon.
In the following, we refer to those collisions as low-energy heavy-ion collisions.
Fusion is not necessarily achieved in the low-energy heavy-ion collisions, neither is fragmentation. 
Reaction dynamics and the resulting products can be drastically different depending on the impact parameter, the mass of colliding ions, and so on.
Therefore, the reaction mechanism of low-energy heavy-ion collisions is worth investigating to understand the production of chemical elements.
This is deeply concerned with an open problem as for the existence and the origin of chemical elements including their production mechanism.  

Charge equilibrium in heavy-ion reactions means the states with the proton-neutron ratio corresponding to the average of the two colliding ions, and charge equilibration is the process leading towards charge equilibrium.
The chemical property of final products can be different depending on whether charge equilibration appears or not, being determined by the number of protons.
There is a relatively long research history for the charge equilibration.
In fact many experiments of low-energy heavy-ion collisions were carried out in the 1960's.  
In those experiments most of the final products were reported to be in charge equilibrium, even when the fragmentation takes place~\cite{FREIESLEBEN}.
There should not be any restrictions to the proton-neutron ratio of the final products if the final product forms a kind of stable bound system, so that these experimental results cannot be trivially understood.

One of the most important features of charge equilibration is its rapidness taking as much as a few 10$^{-22}$~s~\cite{FREIESLEBEN}.
This time scale is actually short in the order of magnitude compared with the typical reaction time of low-energy heavy-ion collisions (1000~fm/c~$\sim$~10$^{-20}$~s), so that charge equilibration has been recognized as an inevitable and dominant process in low-energy heavy-ion reactions. 
The relation between charge equilibration and the isovector giant dipole resonance has been studied relatively well because of the correspondence in their time scales, but no decisive conclusion has ever been obtained.
Indeed, including the question of ``when does charge equilibration take place ?'', many things could not be explained merely by the isovector giant dipole resonance. 
For the theoretical research on the relation between charge equilibration and the isovector giant dipole resonance, the importance of dipole mode to charge distribution of a fissioning nucleus was pointed out by Hill-Wheeler~\cite{HILL} in the 1950's.
Research on charge equilibration using time-dependent mean field calculations were started in the 1970's (Bonche-Ng\^o~\cite{BONCHE}).
Recently research based on three-dimensional time-dependent mean field calculations was carried out by Simenel-Chomaz-de France~\cite{SIMENEL,IWATA}.
In this article, based on Ref.~\cite{IWATA}, the unknown relation between charge equilibration and the zero sound propagation is presented.
It leads to a rather universal recognition of zero sound propagation in femto-scale quantum liquids, which cannot necessarily be reduced to the giant dipole resonance.  

\section{Theory of matter wave propagation}
\subsection{Landau's Fermi liquid theory}
Let us denote a many-particle wave function by $\Psi(t,x_1,\cdots,x_n)$.
It is assumed to satisfy
\[ 
\displaystyle i \hbar \frac{\partial \Psi(t,x_1,\cdots,x_n) }{\partial t} = H \Psi(t,x_1,\cdots,x_n), 
 \]
where $H$ denotes the Hamiltonian operator.
The solution can be represented by $\Psi(t,x)=e^{-itH/\hbar} \Psi(0,x)$ under a suitable boundary condition, if $H$ does not depend on $\Psi(t,x_1,\cdots,x_n)$.
Let the corresponding probability density be $\rho(t,x)$, and begin with the classic theory of sound propagation inside gases.
Readers may wonder why sound propagation is related to charge equilibration, it will be clarified step by step.
Let the equilibrium probability density be $\rho_0(x)$.
If the fluctuation is added to the equilibrium:
\[ \rho (t,x) = \rho_0(x) + \delta \rho (t,x), \]
a force arises from the gradient pressure.
Here the essential property of sound propagation is extracted from the simplified linearized analysis.
The equation of motion is given by
\begin{equation} \label{eq2-1}  \frac{\partial (\rho {\bf v}) }{\partial t} \sim \rho_0 \frac{\partial {\bf v}} {\partial t} = -  \nabla P, \end{equation}
where ${\bf v}$ and $P$ denote the velocity and the pressure, respectively. 
The pressure $P$ depends on both $\rho$ and the entropy $S$. 
On the other hand, the equation of continuity is given by
\[  \frac{\partial \delta \rho }{\partial t} = - div (\rho {\bf v}). \]
The right hand side is approximated by $- \rho_0  ~ div~{\bf v}$ to the lowest order. 
After differentiating this equation by $t$, an equality is derived together with the divergence of Eq. (\ref{eq2-1}).
If we further assume that $P$ is expanded with respect to $\delta \rho$ at a fixed entropy, the lowest order contribution brings about
\begin{equation}  \frac{\partial^2 \delta \rho }{\partial t^2} =  \left( \frac{\partial P}{\partial \rho} \right)_S  \Delta \delta \rho.  \end{equation}
This is a wave equation for $\delta \rho$, where $ \sqrt{ \left( \partial P/\partial \rho \right)_S }$ represents the propagation speed of the density. 
For example, $ \sqrt{ \left( \partial P/\partial \rho \right)_S }$ is given by $v_F/\sqrt{3}$ for the perfect Fermi gas in its ground state ($S=0$), where $v_F$ means the Fermi velocity.
This type of propagation is called first sound, which provides a picture for particles propagating with changing density.

On the other hand, we should pay attention to the propagation of particle without changing density.
This type of propagation is called zero sound, which has a finite frequency even when the wave number is equal to zero (cf. zero point vibration).
To understand zero sound, the linear response is considered for a given external field $H_{ex}(t)$:
\[ \left\{ \begin{array}{ll} 
\displaystyle i \hbar \frac{\partial \Psi(t,x_1,\cdots,x_n) }{\partial t} 
= (H + H_{ex} (t) )  \Psi(t,x_1,\cdots,x_n), \vspace{1.5mm} \\
\displaystyle H_{ex} (t) = \int  d^3 x ~ n(t,x) ~ U_{ex}(t,x),
\end{array} \right. \]
where $n(t,x)$ denotes the number density operator.
In particular, if we restrict ourselves to the impulsive perturbation $ U_{ex}(t,x) =  {\bar U}_{ex} e^{iq x} \delta(t)$, the linear response is
\[ \begin{array}{ll} 
\delta < n(t,x) >  \vspace{1.5mm} \\
 = {\bar U}_{ex} e^{iq \omega} (2 \pi)^{-1} \int~ d\omega ~ e^{-i\omega t} {\bar U}_{ex}^{-1}(q) \left( [ \kappa^R (q,\omega) ]^{-1} -1 \right), 
\end{array}  \]
where $\kappa^R (q,\omega)$ denotes the retarded generalized dielectric function.
The resonance frequency can be calculated by the pole of the integrand of the right hand side.
Eventually we assume the phonon dispersion relation ($\Omega_q = c_0 q$), which reproduces the frequency of propagating wave, and is nonzero even when the wave number is equal to zero:
\[ 1 = V(q) \Pi^{0~R} (q, \Omega_q -i \gamma_q),  \]
then the resonance frequency is obtained.
Consequently the dispersion relation at long wave length becomes
\[ \lim_{q \to 0} \frac{\pi^2 \hbar^2}{m k_F V(q)} = \frac{1}{2}x \log \left( \frac{x+1}{x-1} \right) -1, \]
where $x = m \omega / \hbar k_F q$; $m$ and $k_F$ denote the mass and the Fermi wave number, respectively.
Note that this relation is the representation in momentum space. 
The lower-limit of the velocity for zero sound is the Fermi velocity, because the non-damping mode can only exist when $x>1$ (see the denominator inside the logarithmic function).
If $V(q) \to V(0)$ is assumed in the limit $q \to 0$, the left hand side becomes $\pi^2 \hbar^2/(m_F V(0))$, and the propagation velocity of zero sound is represented using $V(0)$.
For the weak-coupling limit ($V(0) << \hbar^2/m k_F$):
\[c_0 \sim v_F \left\{ 1 + 2 \exp \left( - \frac{2 \pi^2 \hbar^2}{m k_F V(0)} -2  \right) \right\}, \]
and for the strong-coupling limit ($V(0) >> \hbar^2/m k_F$):
\[ c_0 \sim v_F \left\{ \frac{V(0)}{3 \pi^2(\hbar^2/mk_F)}  \right\}^{1/2}.  \]
This shows that the propagation velocity of zero sound is almost equal to $v_F$ in the weak-coupling limit.
It is just $\sqrt{3}$ times faster than the previously seen first sound velocity of the perfect Fermi gas.
A rather general discussion shows that zero sound is faster than first sound \cite{LANDAU-the1,LANDAU-the2}.
Roughly speaking, the matter waves without changing density propagate more easily than those with changing density, because the change in density possibly leads to the appearance of larger restoring force.
Zero sound provides a picture of particle exchange within a quite short time (but not instantaneous).

Two different sound propagations in Fermi liquids have been discussed in terms of whether they entail the density change or not.
In particular there exists zero sound in Fermionic many-body system, which is different from and faster than the ordinary sound.
Zero sound has been known to arise from the collective dynamics of the Fermionic many-body systems.
All the collective dynamics is actually based on the propagation of zero sound, while there are various representations (various modes) for the collective dynamics.  
One distinct difference between zero and first sounds is their relation to collisions between particles.
First sound appears when the states are in local thermal equilibrium.
This corresponds to the situation when the mean interparticle collision time is sufficiently smaller than the oscillation period of the propagating wave.
Meanwhile zero sound is associated with the collective excitation mode, which disappears when there are many collisions between particles.
This corresponds to the situation when the mean interparticle collision time is larger than the oscillation period of the propagating wave.
Although such a collisionless energy regime is expected to be realized due to the Pauli principle, its validity should be confirmed for an individual physical system.
Consequently zero sound is expected to be important to low-energy phenomena, while first sound becomes effective at higher energies. 
For the usage of the terminology low and high energies, we had better note that both sounds are scaled by $v_F$ (i.e. Fermi energy). 
For more details of sound propagation, see Sec. 5 of Ref.~\cite{FETTER}.

\subsection{Nucleon propagation in heavy-ion collisions}
As a theoretical research on the zero sound in Fermionic many-body systems, Landau's Fermi liquid theory~\cite{LANDAU57} is well known, where zero sound propagation was actually seen in liquid $^3$He~\cite{LEGGET}.
However, there is no guarantee that such a sound plays a role in heavy-ion reactions.
There are two essential differences between heavy-ion reactions and the $^{3}$He case; the physical system consists of finite numbers of nucleons, and the event is accomplished within a finite time interval.
That is to say, both size and time are highly restricted in heavy-ion reactions.
Indeed, in the context of many-nucleon systems, the main interest of zero sound propagation was not in heavy-ion reactions but in nuclear vibrations (for example, see Ref.~\cite{RING}). 
In the following the mechanism of charge equilibration is discussed with respect to whether it is achieved by nucleon propagation with or without changing density.

\section{Mechanism of charge equilibration}

\begin{figure*}
\begin{center}
\includegraphics[width=17.5cm]{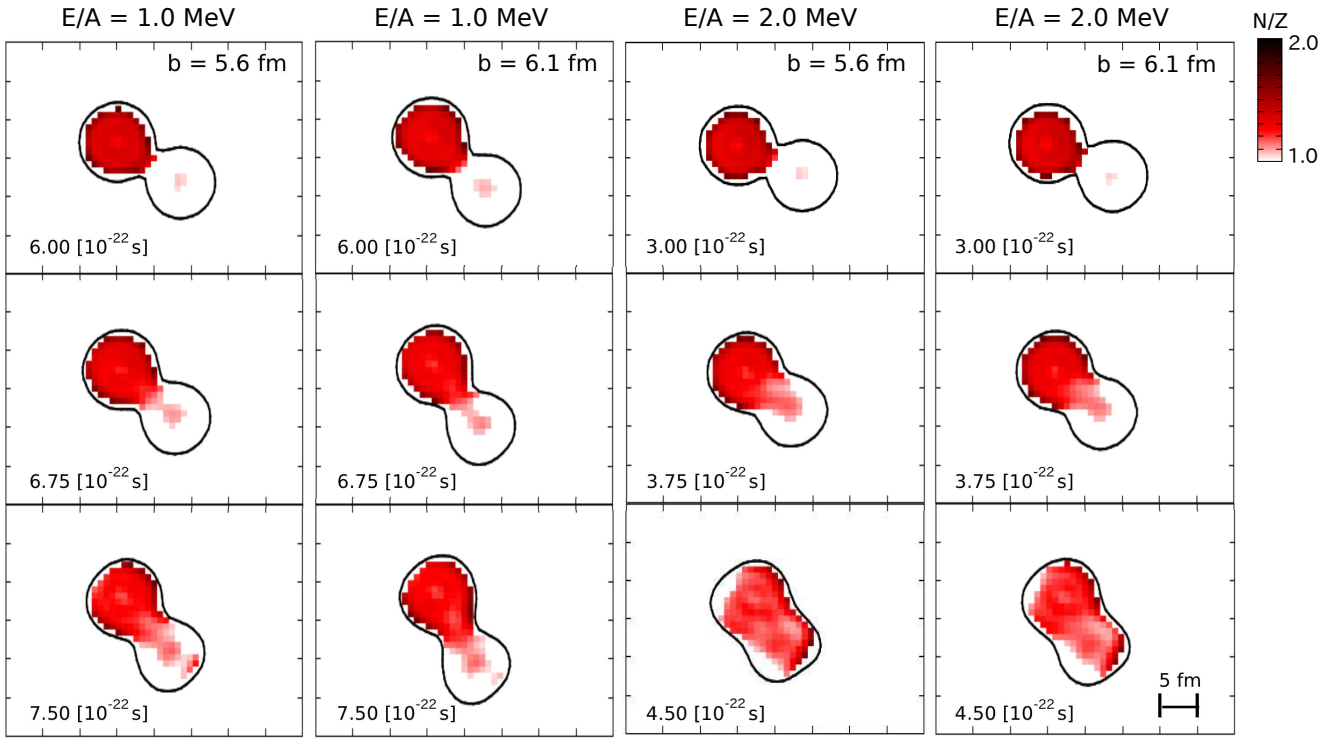} 
\caption{\label{fig1} (Color online) Propagation of neutron-rich flow is depicted for the collision between $^{52}$Ca and $^{36}$Ca, where $^{52}$Ca and $^{36}$Ca correspond to the ions coming from the left and right hand sides, respectively.
The colored parts correspond to the parts with $N/Z>1$ (each frame is 40$\times$30~fm$^{2}$), and the density contour equal to 0.02~fm$^{-3}$ is shown by a thick black curve. 
Three-dimensional time-dependent Hartree-Fock calculations with a Skyrme interaction (SLy4d) is carried out; the single-particle wave functions are represented on a Cartesian grid with the spacing of 0.8~fm, and the time unit of calculation is set to 1.5 $\times 10^{-24}$~s.
The initial distance between the two colliding ions are set to 20~fm, then the relative velocity of collision is given. }
\end{center}
\end{figure*}

\subsection{Nucleon propagation realizing charge equilibration}
As an example of Fermionic many-body systems, we consider a many-nucleon system.
There are two types of nucleons, that is, protons and neutrons.
Charge equilibration is the mixing of protons and neutrons due to the time evolution, therefore it is a kind of chemical mixing.
Apart from such a chemical equilibration, there are several kinds of equilibration in heavy-ion reactions, that is, mass equilibration, momentum equilibration, and thermal equilibration.   
Among them charge equilibration (chemical equilibration) has drawn special attention because of its crucial role in the synthesis of chemical elements. 

Let us consider the collision of two ions.
The existence of an upper energy limit for charge equilibration has been pointed out by Iwata-Otsuka-Maruhn-Itagaki~\cite{IWATA} for the first time.
The limit is presented by a formula explaining both experiments and numerical calculations based on microscopic three-dimensional time-dependent mean-field theory.
Furthermore, the upper energy limit has been concluded to be determined by the Fermi energy.
\begin{equation} \begin{array}{ll} 
\label{eq1}
\displaystyle \frac{E_{lab}}{A}  
= \frac{\hbar^2 (3 \pi^2 \rho_{\rm min})^{2/3}}{2m} 
+ \frac{e^2 Z_1 Z_2}{4 \pi \epsilon_0 r_0} \frac{A_1 + A_2} {A_1 A_2 (A_1^{1/3}+A_2^{1/3})}, 
\end{array} \end{equation}
\begin{eqnarray} 
\label{eq1d}
 &\rho_{\rm min} = {\displaystyle \min_{i}} \left( \frac{N_{i}  \left( \frac{4 \pi r_0}{3}  A_i^{1/3} \right)^{-1} }{(1- 3{\bar \epsilon})(1+{\bar \delta})}, \frac{Z_{i}  \left( \frac{4 \pi r_0 }{3} A_i^{1/3}\right)^{-1} }{(1- 3{\bar \epsilon})(1-{\bar \delta})}\right), 
\end{eqnarray}
where $i= 1,2$ identifies the two colliding ions, and $E_{lab}$ means the energy in the laboratory frame.
$A_1$ and $A_2$, which satisfy $A= A_1+A_2$, $A_1 = Z_1 + N_1$ and $A_2 = Z_2 + N_2$, denote the masses of the two colliding ions, where $Z_1$, $Z_2$, $N_1$ and $N_2$ denote the proton and the neutron numbers of each colliding ion (labeled by $i$), respectively. 
 ${\bar \epsilon}$ and ${\bar \delta}$ are parameters introduced based on Ref.~\cite{MYERS}. 
This formula provides an upper energy limit of charge equilibration, which arises from the nucleons propagating at the Fermi velocity.
In heavy-ion collisions, there are four different Fermi velocities (Fermi energies), because there are two different kinds of nucleons and two colliding ions.
The first term on the right hand side of Eq. (\ref{eq1}) calculates those four Fermi velocities.
In particular, for the many-nucleon system, special structures such as skin and halo exist, and those structures change the Fermi energy.
These effects are taken into account in ${\bar \epsilon}$ and ${\bar \delta}$ in Eq. (\ref{eq1d}).
For the nucleon wave associated with each Fermi energy to propagate throughout both colliding ions after touching, the minimum is taken in Eq. (\ref{eq1d}).
This treatment does not seem to be so important at a glance, but it contributes to derive the fact that different reactions having exactly the same composite nucleus can result in a different upper energy limit (for the difference of upper energy limit, see Fig.2 of Ref.~\cite{IWATA}).
If it is not for such a treatment, we cannot explain both numerical calculations and experiments.
Furthermore, for a correct comparison, it is necessary to estimate the relative velocity at contact time.
The most considerable effect here is the deceleration due to the Coulomb repulsion, which becomes prominent for cases when the masses of the colliding ions are larger.
This effect is considered in the second term on the right hand side of Eq.(\ref{eq1}).
Consequently it was confirmed in Ref.~\cite{IWATA} that the final fragments achieving charge equilibrium drastically decrease at an incident energy higher than the energy shown in the formula. 

The nucleons propagating at the Fermi velocity, which is represented by Eqs. (\ref{eq1}) and (\ref{eq1d}), correspond to the propagation of zero sound.
Indeed, according to the calculation of the mean free path of nucleons inside nuclear matter~\cite{COLLINS}, collisions between nucleons were shown to appear and increase rapidly if the incident energy of heavy-ion collisions becomes a few 10 \% larger than the Fermi energy (of the composite nucleus).
In such situations, zero sound itself disappears. 
The possible propagation speed of zero sound is not not different from the amplitude of the Fermi velocity.
This fact explains the reason why Eqs. (\ref{eq1}) and (\ref{eq1d} are related to zero sound propagation.  

Figure~\ref{fig1} visualizes the propagation of charge equilibrating flow: neutron-rich flow from $^{52}$Ca (N/Z = 32/20) to $^{36}$Ca (N/Z = 16/20).
Note here that, because this heavy-ion reaction is classified to class II of the classification shown in Fig. 1 of \cite{IWATA-np} and the ``N/Z= 1''-line is located between the initial points of  $^{52}$Ca and $^{36}$Ca on the N-Z plane, its charge equilibration dynamics can be measured only by the flow of neutrons.
Cases with two different impact parameters and two different energies are compared.
The final product are different, where only the case with $b = 6.1$~fm and $E/A=$1.0~MeV results in fragmentation, while the other cases result in fusion.  
Despite such significant differences in final products, the propagation speed of charge equilibrating flow is almost the same.
It can be confirmed by the time evolutions of 0.75$\times$10$^{-22}$~s after the neck formation (the neck is formed at 6.0$\times$10$^{-22}$~s for $E/A=$1.0~MeV cases and 3.0$\times$10$^{-22}$~s for $E/A=$2.0~MeV cases).
It shows that the propagation speed of charge equilibrating flow is mostly independent of the incident energy (the relative velocity of collision) and the impact parameter, while the total contained neutrons in neutron-rich flow depends highly on the incident energy. 
In particular the propagation speed of charge equilibrating flow is faster than the relative velocity of collision (Table~\ref{tab1}).

\begin{table*}
\caption{\label{tab1}  Comparison of speeds, where $|v_F|$ is fixed to 1/3 of the speed of light (corresponding to the nuclear standard value). The propagation speed of charge-equilibrating flow is calculated by the propagation speed of the wave front of $N/Z = 1.10$.
The relative velocity of collision at the contact is slower than that at the initial time, because of the deceleration due to the Coulomb repulsion. } 
\begin{tabular}{|l|c|c|} \hline
Motion & Speed & Description \\ \hline
Propagation of charge-equilibrating flow \quad&\quad  0.90~$|v_F|$ \quad&\quad $\sim$ 6.5/(0.75$\times 10^{-22}$)~fm/s \quad \\ \hline
Relative velocity for $E/A$=2.0 MeV \quad&\quad  0.36~$|v_F|$ \quad&\quad Speed given at the initial time \quad \\ \hline
Relative velocity for $E/A$=1.0 MeV \quad&\quad  0.23~$|v_F|$ \quad&\quad Speed given at the initial time \quad \\ \hline
\end{tabular}
\end{table*}

With respect to zero sound, what was clarified in Ref.~\cite{IWATA} can be summarized in the following three points.
First, zero sound propagation plays a role in heavy-ion collisions.
Second, the fast charge equilibration, which is achieved within the order of $10^{-22}$~s, is realized as nucleon propagation without changing density.
Third, there exists an upper energy limit for the fast charge equilibration, which corresponds to the energy limit at which zero sound can survive. 
For the terminology of ``fast'' charge equilibration, it takes into account the existence of another kind of charge equilibration that has nothing to do with zero sound propagation.
Such charge equilibration, which appears at higher energies, is more related to the first sound.
Therefore its process is relatively slow compared to the fast charge equilibration, and insufficient to lead to fusion or to charge equilibrium for most fragments.
Charge equilibration at higher energies was also studied well (for example, see references [8-10, 20-23] of Ref.~\cite{IWATA}).

As a remark, the previous research on charge equilibration with respect to the collective dynamics is mentioned.
As is discussed, charge equilibration was studied in association with the isovector giant dipole resonance.
Charge equilibration is sometimes related to the isovector dipole resonance, but not in all cases; i.e. the concept of resonance is too restrictive to explain charge equilibration.
Nevertheless isovector giant dipole resonance means that the modes related to the composite nucleus play a role, it cannot explain the different upper energy limits for different reactions having exactly the same composite nucleus.
In addition, an isovector mode different from the isovector giant dipole resonance sometimes appears (see Fig. 4 of Ref.~\cite{IWATA}).
Although it is always true that the fast charge equilibration is achieved by the collective dynamics, the fast charge equilibration is not necessarily achieved by the isovector giant dipole resonance. 

\subsection{Origin of charge equilibration}
Apart from how and when charge equilibration takes place, here we see the reason why charge equilibration takes place.
First of all, the propagation of zero sound is expected to be efficient to any kind of nucleon propagations and vibrations.
The answer is obtained by clarifying the origin of charge equilibration.
When the two ions have a contact during the heavy-ion collisions, large fluctuations appear in the shape and the internal structure.
Zero sound is expected to contribute to stabilization by changing both the shape and the internal structure, which can be understood by the contribution of each term included in the Bethe-Weizs\"acker mass formula~\cite{BETHE}:
\[ B(A) = a_{\rm vol}A + a_{\rm surf} A^{2/3} + a_{\rm coul} Z^2 A^{-1/3} + a_{\rm sym} \frac{(N-Z)^2}{A}, \]
where the coefficients are given by $a_{\rm vol} \sim -16$~MeV, $a_{\rm surf} \sim 20$~MeV, $a_{\rm coul} \sim 0.751$~MeV and $a_{\rm sym} \sim 21.4$~MeV, respectively~\cite{GREINER}.
After contact, the first and second terms contribute to stabilization by changing the shape such as the volume and surface area, while the third and forth terms contribute mainly to stabilization by changing the internal structure.
It is reasonable that the shape change (including vibration) due to zero sound propagation leads to stabilization, and we do not go further into detail.
Here we focus on stabilization by changing the internal structure.
The effect due to the symmetry energy (the forth term) should play a principal role, because the Coulomb energy (the third term) is actually small except for the collisions between very heavy ions.
This symmetry energy is the principal driving force of the stabilization by changing the internal structure, and its effect acting on zero sound propagation is fast charge equilibration.

\section{Summary}
The propagation of charge equilibrating flow has been visualized, and a physical interpretation has been given to the fast charge equilibration; i.e., the fast charge equilibration is realized by zero sound propagation.
This means that the collective dynamics of the many-nucleon system and thus its Fermionic statistical properties are essential to the fast charge equilibration.
In this context the upper energy limit for the fast charge equilibration corresponds to the limit energy for zero sound propagation to be effective.
Consequently charge equilibration should be regarded as the exchange of nucleons (charge exchange), because zero sound does not necessarily entail density change. 

As is discussed, fast charge equilibration is not a process appearing only at a certain energy.
Fast charge equilibration universally appears in low-energy heavy-ion reactions at energies lower than the upper energy limit, instead.
Here note that the fast charge equilibration process becomes operational only if the two colliding ions interact by the nuclear force.

The correspondence of zero sound propagation in femto-scale systems is not only the giant dipole resonance, but also the flow propagating at the Fermi velocity.
Indeed, the similarity between neutron-rich flow shown in Fig. \ref{fig1} and the isovector giant dipole resonance is quite limited.  
It is suggested that a prominent role of zero sound in femto-scale systems can be experimentally detected by the appearance of the fast charge equilibration; whether most of the final products of heavy-ion reactions are in charge equilibrium of not. \\

This work is based on the collaboration with  Profs. T. Otsuka, J. A. Maruhn, and N. Itagaki, and supported by the Helmholtz alliance HA216/EMMI. 
The author thanks to Prof. Maruhn for reading the manuscript carefully.
Encouragement from Dr. C. Simenel at the initial stage of this research is acknowledged.

\end{document}